\documentstyle[namedreferences,epsf,NATO]{crckapb} 
\def\lesssim{\mathrel{\hbox{\rlap{\hbox{\lower4pt\hbox{$\sim$}}}\hbox{$<$}}}}

\begin{opening}
\title{THE  $h - \Omega_{o}\;\:\:$ DIAGRAM FROM RECENT 
CMB OBSERVATIONS}
\subtitle{YES, WE CAN ALREADY SAY SOMETHING}

\author{C.H. LINEWEAVER}
\institute{Observatoire de Strasbourg\\
           Strasbourg, France\\
		and\\
           School of Physics, UNSW\\
           Sydney, Australia\\
           charley@edwin.phys.unsw.edu.au}
\end{opening}

\runningtitle{THE $h-\Omega_{o}$ DIAGRAM}

\def\thefirstfig{
\makebox{
\medskip
\noindent
\parbox[l]{1.5truein}{
\footnotesize
{\bf Figure 1.\\ Current CMB Data}
The solid line is the best-fitting model in the
open and critical density family of models.
The $\chi^{2}$ fit is good.
The grey band is the variation of these
models permitted within the 68\% confidence region of
Figure 4.
Experiments designed optimally to constrain parameters should
have window functions in regions where the vertical spread 
of the grey band is largest.
Figure from Lineweaver \& Barbosa (1997b).}
\hglue0.1truein
\parbox[r]{3.3truein}{\epsfxsize=3.3truein\epsfbox{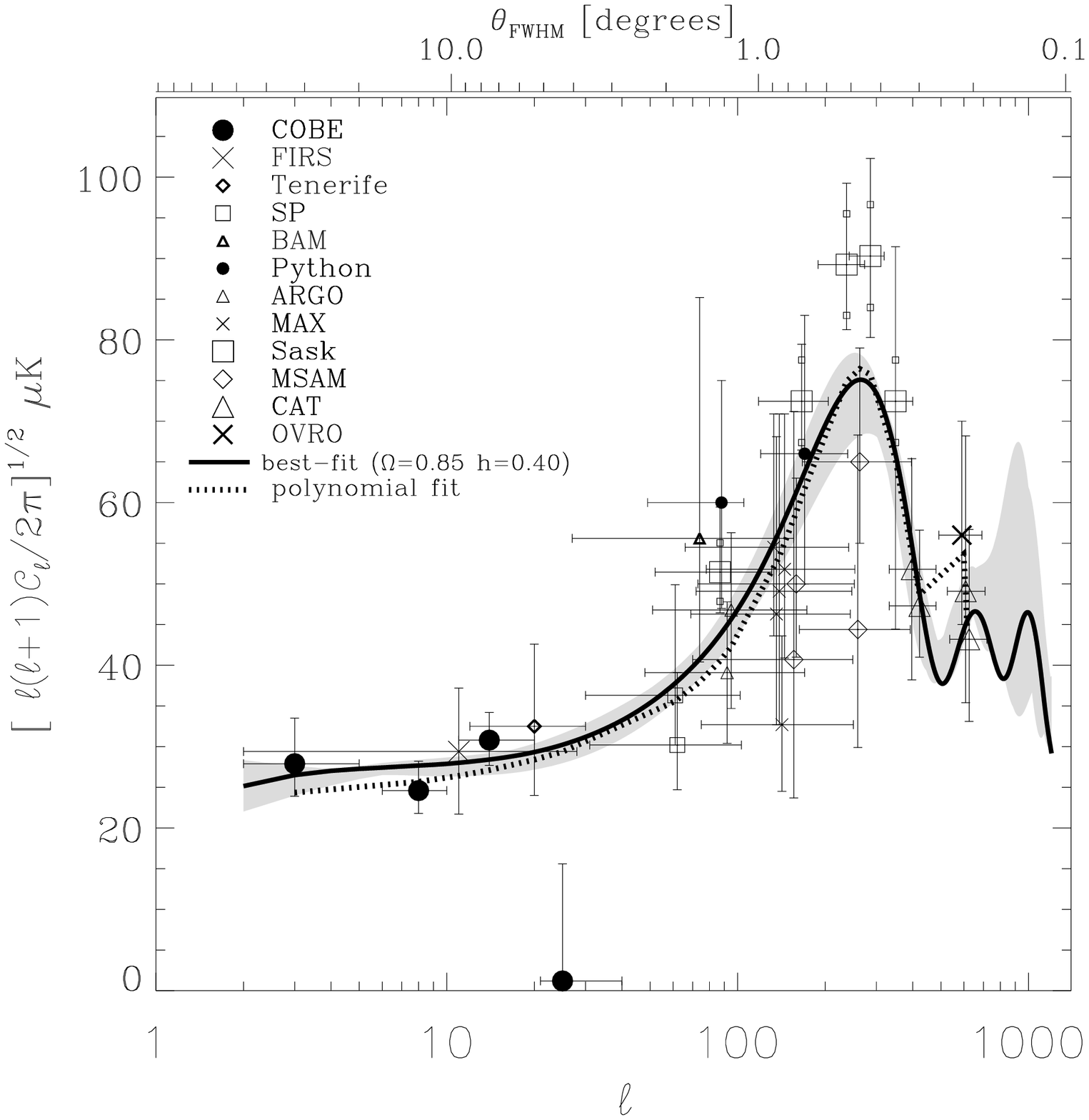}}
\smallskip
}
}

\def\thesecondfig{
\makebox{
\medskip
\noindent
\parbox[l]{1.7truein}{\footnotesize
{\bf Figure 2.\\  CMB Constraints in the $h - \Omega_{b}$ plane 
when $\Omega_{o} = 1$.}
The dark grey area is the approximate $68\%$ confidence region
about the best-fit at $h=0.30^{+0.18}_{-0.07}$
(marked with a {\bf X}).
The light grey band is from big bang nucleosynthesis 
($ 0.010 < \Omega_{b}\:h^{2} < 0.026$). 
Figure from Lineweaver \& Barbosa (1997a).}
\hglue0.1truein
\parbox[r]{2.9truein}{\epsfxsize=2.9truein\epsfbox{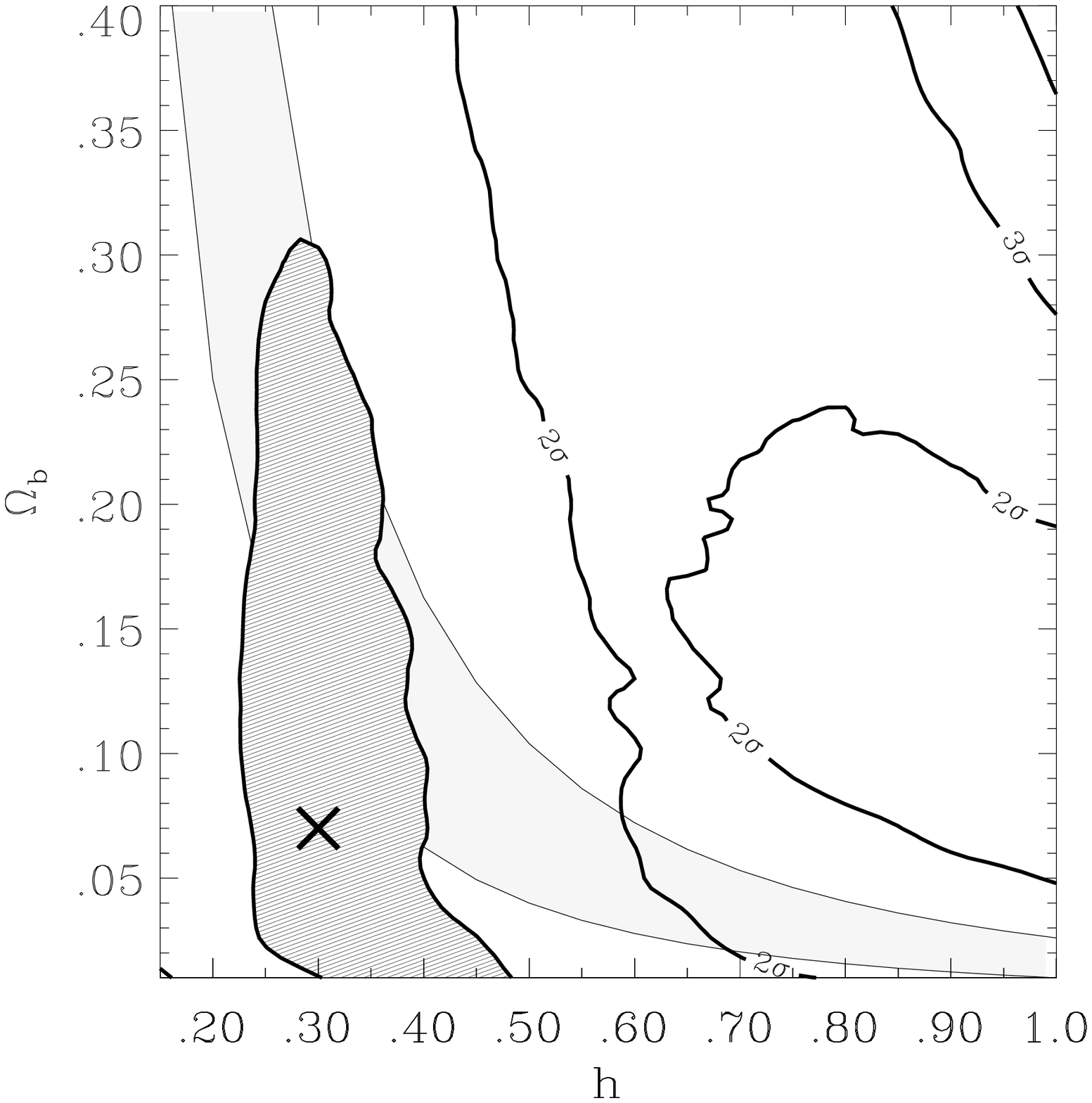}}
\smallskip
}
}
\def\thethirdfig{
\makebox{
\medskip
\noindent
\parbox[l]{1.7truein}{\footnotesize
{\bf Figure 3.\\  Non-CMB constraints in the $h-\Omega_{b}$ plane
when $\Omega_{o} = 1$}
The light grey band is from big bang nucleosynthesis
and is the same in Figure 2. 
The two other bands are the preferred regions from 
baryonic fractions in clusters and the matter power spectrum ($\Gamma$).
The vertical lines indicate the range for the age of the Universe
inferred from the oldest stars in globular clusters.
An approximate joint likelihood of these constraints
yields the dark grey $68\%$ confidence region:
$h=0.40 \pm 0.07$.
Figure from Lineweaver \& Barbosa (1997a).}
\hglue0.1truein
\parbox[r]{2.9truein}{\epsfxsize=2.9truein\epsfbox{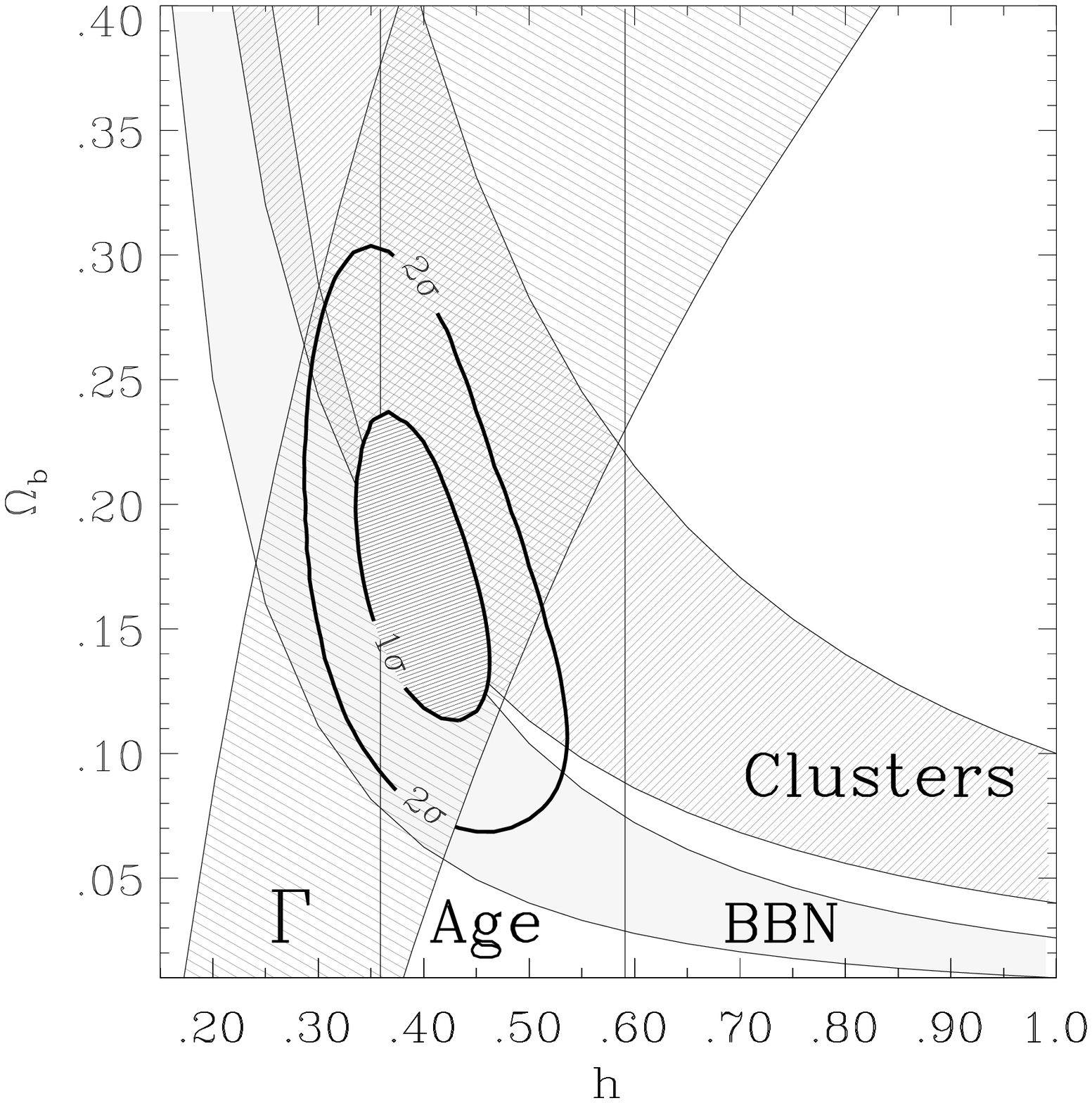}}
\smallskip
}
}

\def\thefourthfig{
\makebox{
\medskip
\noindent
\parbox[l]{1.7truein}{\footnotesize
{\bf Figure 4.\\ CMB constraints in the $h-\Omega_{o}$ plane.}
The dark grey region is the 68\% confidence region
around the best-fit at $h=0.40^{+0.57}_{-0.14}$
(marked with an {\bf X}).
The narrow banana-shaped region means that plausible values of
$h$ and $\Omega_{o}$ are correlated and this region defines a new constraint:
$\Omega_{o} h^{1/2} = 0.55 \pm 0.10$.
The age constraint shown is $10 < t_{o} < 18$ Gyr.
Figure adapted from Lineweaver \& Barbosa (1997b).}
\hglue0.1truein
\parbox[r]{3.1truein}{\epsfxsize=3.1truein\epsfbox{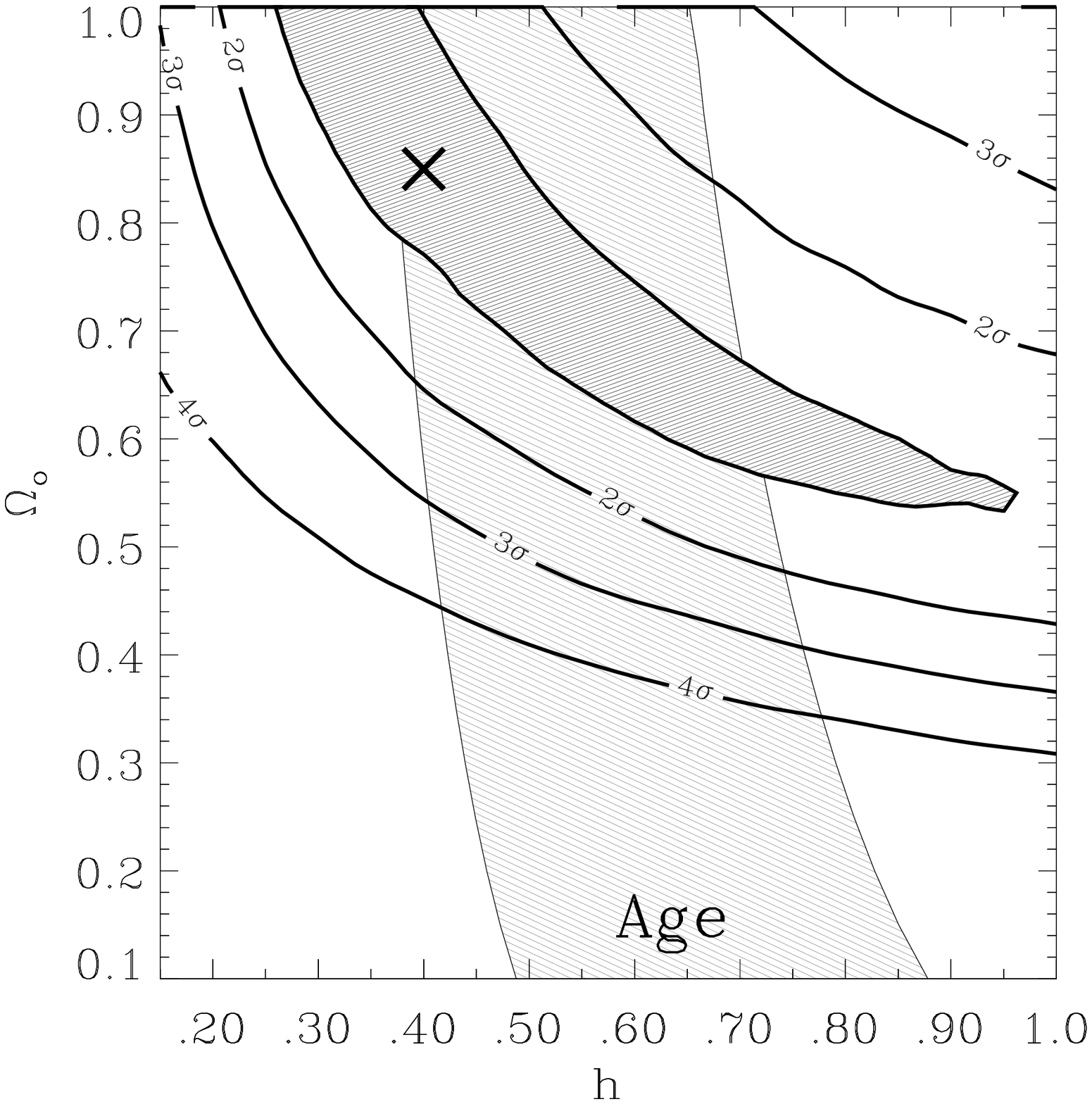}}
\smallskip
}
}

\def\thefifthfig{
\makebox{
\medskip
\noindent
\parbox[l]{1.7truein}{\footnotesize
{\bf Figure 5.\\ Non-CMB constraints in the $h - \Omega_{o}$ plane.}
Approximate joint likelihood from the baryonic 
fraction in clusters ($0.10 < \Omega_{o}\;h^{1/2} < 0.65$), 
globular cluster ages ($10 < t_{o} < 18$ Gyr), 
the matter power spectrum ($0.169 < \Gamma < 0.372$) and
local measurements of Hubble's constant ($0.50 < h < 0.80$).
Figure from Lineweaver \& Barbosa (1997b).}
\hglue0.1truein
\parbox[r]{3.1truein}{\epsfxsize=3.1truein\epsfbox{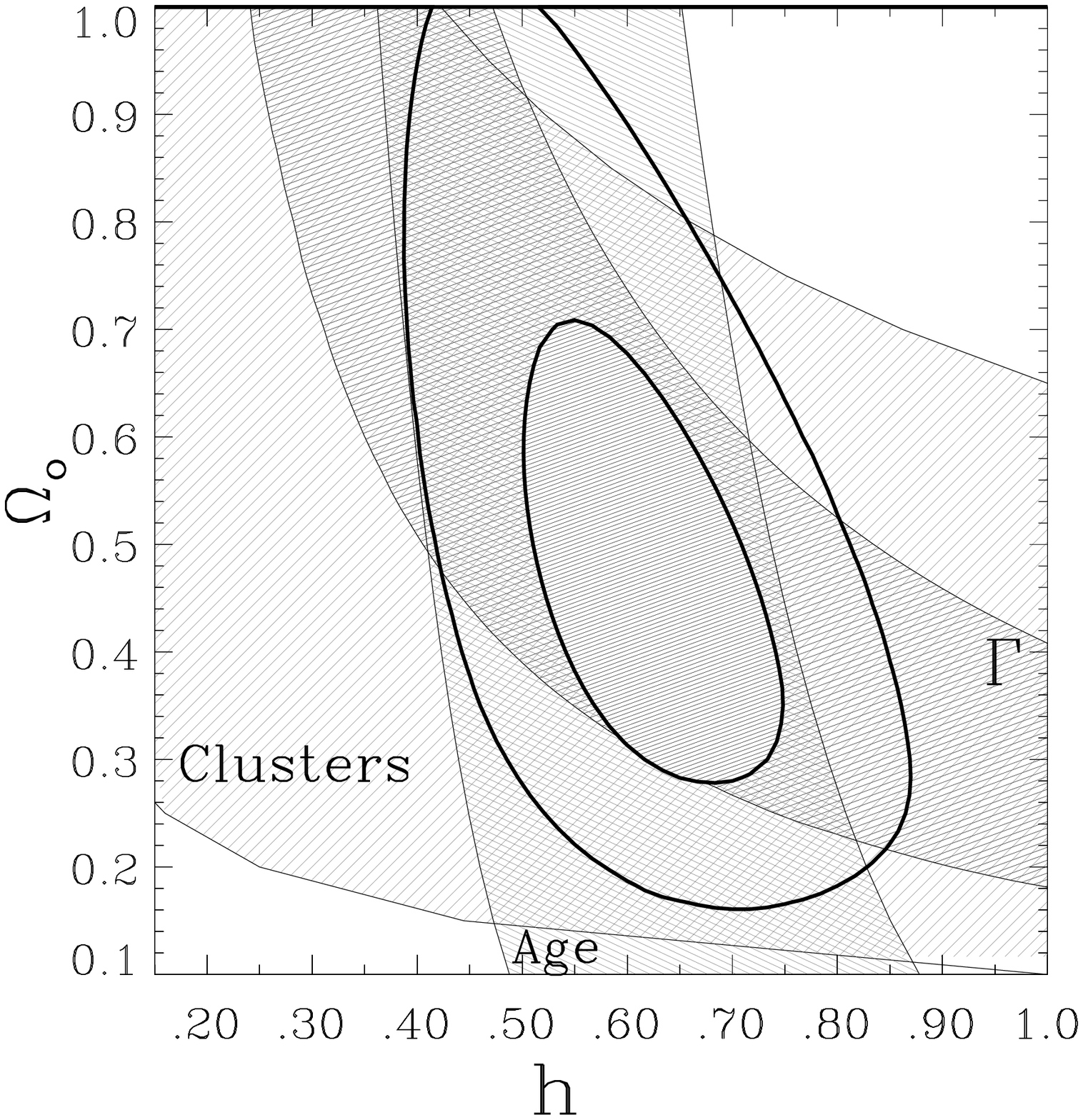}}
\smallskip
}
}

\def\thesixthfig{
\makebox{
\medskip
\noindent
\parbox[l]{1.7truein}{\footnotesize
{\bf Figure 6.\\ CMB + non-CMB constraints in the $h - \Omega_{o}$ plane.}
When the CMB constraint from Figure 4 ($\Omega_{o}h^{1/2}=0.55\pm 0.10$)
is combined with the non-CMB constraints in Figure 5 we obtain 
$h=0.58\pm 0.11$ and $\Omega_{o}=0.65^{+0.16}_{-0.15}$.

Figure from Lineweaver \& Barbosa (1997b).}
\hglue0.1truein
\parbox[r]{3.1truein}{\epsfxsize=3.1truein\epsfbox{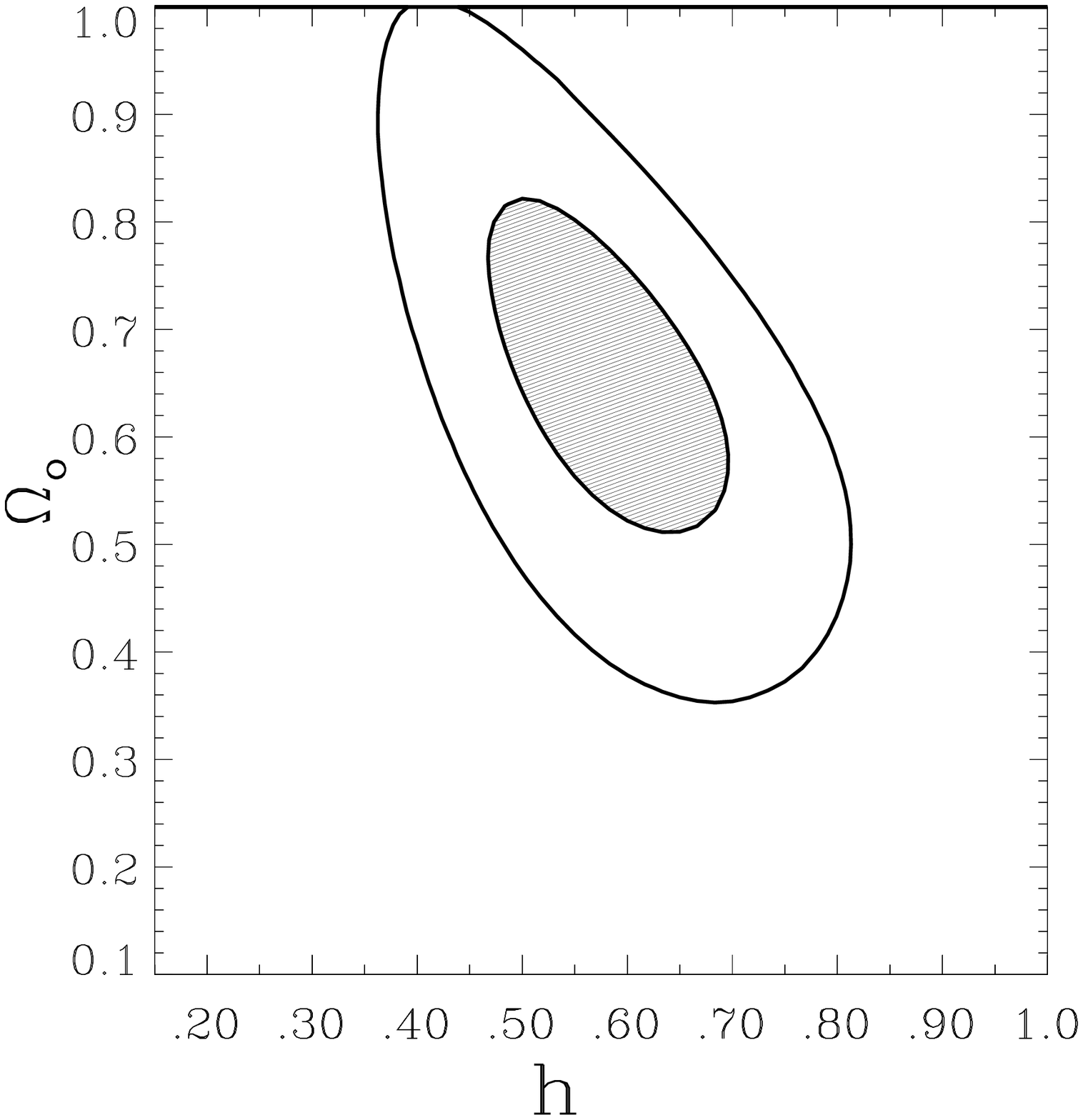}}
\smallskip
}
}

\begin{document}
% The \begin{document} command comes after the \end{opening}
% command.

\begin{abstract}

The CMB is already one of the pillars of the Big Bang model.
However it may also become our most powerful tool to
distinguish contending models and to determine their
cosmological parameters.
To realize this goal, more than 20 observational groups and two
new satellites are gearing up to make precise measurements of the 
CMB at small angular scales. In such a situation it is
important to keep track of what the CMB data can already say
about cosmological parameters.

Current CMB data can already be used to constrain cosmological
parameters. The results are model dependent.
We have obtained contraints on Hubble's constant $h$ and the
density of the Universe $\Omega_{o}$ in the context 
of open and critical density CDM models with $\Lambda=0$.
In critical density models we obtain $h=0.30^{+0.18}_{-0.07}$.
This low value is inconsistent with direct measurements of $h$ but
fully consistent with four other cosmological measurements:
Big Bang nucleosynthesis, cluster baryonic fraction, 
age constraints from globular clusters and limits on the shape 
parameter $\Gamma$ of matter power spectra (in $\Omega_{o}=1$ models).
If $\Omega_{o}$ is left as a free parameter the constraints on
$h$ are less restrictive:  $h=0.40^{+0.57}_{-0.14}$.
This is fully consistent with local $h$ measurements
and the four other cosmological measurements  mentioned above.
The best-fit density is $\Omega_{o}= 0.85$ and we set an upper limit of 
$\Omega_{o} > 0.4$ at $\sim 95\%$ CL.

At this conference Ostriker has claimed that open-CDM models with
$\Omega_{o} \approx 0.3$ and  $h \approx 0.70$ are compatible with all 
current data. However our new CMB data analysis rules this
model out at more than $\sim 4 \sigma$.
\end{abstract}
  
\section{Introduction}

At this conference it has been said that determinations of
cosmological parameters from CMB data is something that will
be done in the future--that the scatter in the data does
not allow one to conclude anything yet.
However the results presented here show that
current CMB data can {\it already} rule out large regions of
parameter space under certain popular assumptions, e.g.,
adiabatic initial conditions, no tensor modes,
$\Lambda = 0$.
The details of these assumptions as well as results for the 
slope $n$ and amplitude $Q$ of the primordial power spectrum 
and  the position and amplitude of the peak in the power spectrum
are given in Lineweaver \& Barbosa (1997a \& b).
Here I limit the discussion to the CMB constraints on
$h$ and $\Omega_{o}$.

Figure 1 is a compilation of current CMB data.
Although there is much scatter, simple binning of the data
brings out a very significant peak at $\ell \sim 260$.
We do $\chi^{2}$ fits of model power spectra to this data.
The power spectra are calculated with a fast Boltzmann code
(Seljak \& Zaldarriaga, 1996).
Our results are displayed in Figures 2 and 4 where the $\Delta \chi^{2}$
contours are for $\Delta \chi^{2} = [1,4,9]$ corresponding to
$[68\%, 95\%, 99\%]$ confidence regions. 

\bigskip
\bigskip
\noindent
\thefirstfig
%%%%%%%%%%%%%%%%%%%%%%%%%%%%%%%%%%%%%%%%%%%%%%%%%%%%%%%%%%%%%%%
\newpage
\noindent
\thesecondfig
\noindent
\thethirdfig

By comparing Figures 2 and 3 we see that the CMB and Non-CMB constraints
are consistent in the sense that the 68\% confidence regions overlap.
However the overlap is at $h \approx 0.35$.
This is inconsistent with local measurements of Hubble's constant 
which favor $h\approx 0.65 \pm 0.15$. The assumption that $\Omega_{o} = 1$
has led to this inconsistency. 
Figures 4 and 5 show that this inconsistency is 
removed when we let $\Omega_{o} \le 1$.

%%%%%%%%%%%%%%%%%%%%%%%%%%%%%%%%%%%%%%%%%%%%%%%%%%%%%%%%%%%%
\newpage
\noindent
\thefourthfig
\noindent
\thefifthfig

Figure 4 shows that letting $\Omega_{o}$ be a free parameter loosens
the constraint on $h$ but that a very narrow banana 
in the $h-\Omega_{o}$ plane is preferred. This region is consistent
with the non-CMB constraints as well as with
the local measurements of $h$ since the 68\% confidence
regions of Figures 4 and 5 overlap.
An interesting common feature in both plots is that 
high $\Omega_{o}$ values prefer low $h$ while low $\Omega_{o}$ values
prefer high $h$.
Notice that the $\Omega_{o} \approx 0.3$, $h\approx 0.70$ 
model cited by Ostriker as a model consistent with all observational
data is acceptable to the non-CMB data but
is ruled out by the CMB data at more than $\sim 4 \sigma$.
%%%%%%%%%%%%%%%%%%%%%%%%%%%%%%%%%%%%%%%%%%%%%%%%%%%

Figure 6 combines both CMB and non-CMB observations.
The maximum likelihood is at $h=0.58\pm 0.11$ and 
$\Omega_{o}=0.65^{+0.16}_{-0.15}$.
Critical density universes, i.e., $\Omega_{o} = 1$, 
are marginally excluded at about the 95\% confidence level.
The low density universe mentioned by Ostriker ($h\approx 0.70$,
$\Omega_{o} \approx 0.3$) is excluded at slightly more than the 
95\% confidence level.
My conclusion is that CMB data can {\it already}
be used as a tool to distinguish contending cosmological 
models and determine cosmological parameters.

\bigskip
\noindent
\thesixthfig

\end{document}